\documentclass[prd,showkeys,floatfix,nofootinbib,
               preprint,12pt,
               tightenlines,fleqn]{revtex4} 

\usepackage{amsmath,amssymb,revsymb,graphicx,dcolumn}
\usepackage{leftidx}

\newcommand{\beq}{\begin{equation}}
\newcommand{\eeq}{\end{equation}}
\newcommand{\beqa}{\begin{eqnarray}}
\newcommand{\eeqa}{\end{eqnarray}}
\newcommand{\bsubeqs}{\begin{subequations}}
\newcommand{\esubeqs}{\end{subequations}}

\usepackage{enumerate}

\begin{document}
\title[]
      {Holography versus Correspondence principle: \\ 
      eternal Schwarzschild--anti-de Sitter geometry\vspace*{5mm}}
\author{Slava Emelyanov}
\email{viacheslav.emelyanov@physik.uni-muenchen.de}
\affiliation{Arnold Sommerfeld Center for
Theoretical Physics,\\
Ludwig Maximilian University (LMU),\\
80333 Munich, Germany\\}

\begin{abstract}
\vspace*{2.5mm}\noindent
It is shown that the correspondence principle and the holographic principle are incompatible in the
background of an eternal Schwarzschild--anti-de Sitter geometry. The argument is based
on the observation that algebraic structures of local quantum field and CFT operators are not
equivalent. This implies, in particular, the bulk CFT must be singular near the black-hole horizon.
A CFT Hilbert space representation is elaborated which may correspond to the 
AdS black hole in the dual theory.
\end{abstract}


\date{\today}

\maketitle

\section{Introduction}

The holographic principle states that the physical quantum gravity degrees of freedom are located on a 
co-dimension one hypersurface of a given spacetime~\cite{tHooft,Susskind}. The AdS/CFT 
correspondence~\cite{Maldacena1,Gubser&Klebanov&Polyakov,Witten1} is regarded in particular as a 
concrete realization of this principle. Specifically, the quantum gravity degrees of freedom are identified 
with rays in the conformal field theory (CFT) Hilbert space defined on the boundary $\partial\mathcal{M}$ 
of anti-de Sitter space (AdS). AdS spacetime is denoted as $\mathcal{M}_1$ below. The 
duality is further generalised to all geometries being asymptotically AdS space, among of which is the 
Schwarzschild--AdS black hole denoted as $\mathcal{M}_2$ in the following.

Having a boundary CFT field operator algebra, $\mathcal{O}_{\partial\mathcal{M}}$, one can construct its bulk 
analog $\mathcal{O}_{\mathcal{M}}$~\cite{Balasubramanian&Kraus&Lawrence,
Banks&Douglas&Horowitz&Martinec,Balasubramanian&Kraus&Lawrence&Trivedi}. The bulk CFT algebra
$\mathcal{O}_{\mathcal{M}_2}$ is thus treated as a set of all possible observables (if self-adjoint and gauge 
invariant) in the Schwarzschild--AdS geometry. One may expect that quantum gravity should reduce to 
sufficiently well-established local quantum field theory (LQFT) defined on a fixed geometrical background. 
This is a semi-classical limit of quantum gravity, where gravity is governed by a classical theory, i.e. general 
relativity, and matter fields and perturbations of the metric are quantised. This reduction of quantum 
gravity to quantum field theory in curved (classical) spacetimes is a particular type of the correspondence 
principle.\footnote{Rigorously speaking, it is not a
correspondence between quantum and classical theories, but this is a correspondence between
a strongly fluctuating regime of quantum gravity and its weak-field/low-energy limit, when one is 
allowed to treat geometry as approximately classical, i.e. spacetime is classical up to its perturbations 
which are still quantised~\cite{Giddings1,Giddings2,Giddings}.}

Local quantum field theory in globally hyperbolic spacetime is described by an algebra $\mathcal{A}_{\mathcal{M}}$ 
represented in a certain Hilbert space~\cite{Hollands&Wald,Khavkine&Moretti,Fewster&Verch}. 
According to the correspondence 
principle, one should expect there exists a certain CFT algebra $\mathcal{O}_{\mathcal{M}}$ which is isomorphic 
to $\mathcal{A}_{\mathcal{M}}$ in the semi-classical limit. Specifically, this implies
\beqa
[\hat{O}_{\mathcal{M}}(x),\hat{O}_{\mathcal{M}}(x')] &=& 
[\hat{A}_{\mathcal{M}}(x),\hat{A}_{\mathcal{M}}(x')]
\eeqa
for any $\hat{O}_{\mathcal{M}}(x) \in \mathcal{O}_{\mathcal{M}}$ and
$\hat{A}_{\mathcal{M}}(x) \in \mathcal{A}_{\mathcal{M}}$, where $x,x' \in \mathcal{M}{\cup}\partial\mathcal{M}$.
Moreover, a unitary equivalence of theories should be then expected as well. In other words, the CFT 
Hilbert space representation of $\mathcal{O}_{\mathcal{M}}$ must be unitarily equivalent to that of 
$\mathcal{A}_{\mathcal{M}}$. The isomorphism between these theories is well-established in anti-de Sitter
space~\cite{Bena,Hamilton&Kabat&Lifschytz&Lowe,Rehren}.

A physical Hilbert space representation of field operators corresponds to a choice of a physical vacuum. In the AdS 
black-hole geometry, the Hartle-Hawking state, $|\Omega_2\rangle$, is supposed to be a physical one as being 
non-singular on the black-hole horizons. The regularity of $|\Omega_2\rangle$ allows thus to have a self-consistent 
treatment of quantum gravity in the semi-classical regime. A CFT state,  $|\partial\Omega_2\rangle$, is an ``analog'' 
of the Hartle-Hawking state, but for $\mathcal{O}_{\mathcal{M}_2}$. It is expected to be a pure high-energy state which 
is thermal at the semi-classical approximation~\cite{Banks&Douglas&Horowitz&Martinec}. Roughly speaking, if the 
boundary CFT theory is in $|\partial\Omega_2\rangle$, then the bulk geometry should correspond to $\mathcal{M}_2$ 
according to AdS/CFT. 

A question is posed in this paper whether the correspondence principle and holography are compatible
with each other in the eternal Schwarzschild--AdS geometry. We will show below that there is no algebraic
isomorphism between the operator algebras $\mathcal{A}_{\mathcal{M}_2}$ and $\mathcal{O}_{\mathcal{M}_2}$,
i.e. $\mathcal{A}_{\mathcal{M}_2} \not\cong \mathcal{O}_{\mathcal{M}_2}$. We will also elaborate 
on properties of the Hilbert space constructed on the state $|\partial\Omega_2\rangle$.

The outline of this paper is as follows. In Sec.~\ref{sec:adsbh}, two theories based on
$\mathcal{A}_{\mathcal{M}_2}$ and $\mathcal{O}_{\mathcal{M}_2}$ are separately considered.
In Sec.~\ref{subsec:star-algebra}, we will briefly review local quantum field theory in 
the Schwarzschild--AdS geometry. In Sec.~\ref{subsec:2CFT}, we will revisit a description of the AdS black hole by
two CFT operator algebras represented on a factorised product of two CFT Hilbert spaces. The main point will be that
although $|\partial\Omega_2\rangle$ can be represented as the 
thermal-field double state, this does not imply $|\partial\Omega_2\rangle$ belongs to a factorised product of the Hilbert 
spaces in quantum field theory. In Sec.~\ref{subsec:1CFT}, we will argue that the description should be given by a single CFT algebra represented on the state $|\partial\Omega_2\rangle$.

In Sec.~\ref{sec:ai}, these theories are treated together. In Sec.~\ref{subsec:TTT}, it is pointed out that the
Tomita-Takesaki theorem broadly employed to describe a Schwarzschild-AdS black-hole interior by CFT implies
an extra geometrical construction on the black-hole boundary. In Sec.~\ref{subsec:NGT}, it is shown that the
operator algebras $\mathcal{A}_{\mathcal{M}_2}$ and $\mathcal{O}_{\mathcal{M}_2}$ are not algebraically isomorphic.
This leads to an inconsistency when one simultaneously requires both the correspondence principle and the 
holographic principle in the semi-classical regime. 

In Sec.~\ref{sec:discussion}, we will summarise our results and conclude.

Throughout this paper the fundamental constants are set to unity, $c = G = k_\text{B} = \hbar = 1$.

\section{Algebras $\mathcal{A}_{\mathcal{M}_2}$ and $\mathcal{O}_{\mathcal{M}_2}$ and their representations}
\label{sec:adsbh}

According to (extrapolate) AdS/CFT, quantum field theory in the Schwarzschild--AdS geometry can be constructed
either through a quantisation of a classical field defined inside it or through pulling the CFT operators from its boundary.
In other words, one has at a disposal at least two operator algebras, i.e $\mathcal{A}_{\mathcal{M}_2}$ 
and $\mathcal{O}_{\mathcal{M}_2}$, respectively. One may expect these two theories are equivalent at the semi-classical 
approximation. In this section, we will separately discuss these two theories outside of the AdS black hole.

\subsection{Single $*$-algebra and its representations}
\label{subsec:star-algebra}

A quantum field $\hat{\Phi}(x)$, where $x \in \mathcal{M}$, is an operator-valued distribution. To have well-defined
operators built out of it on a certain Hilbert space, it has to be averaged or smeared out over a set of test functions 
$\{f(x)\}$, i.e. functions being smooth and of a compact support. The field operator algebra $\mathcal{A}_{\mathcal{M}}$ 
is a set composed of an identity operator $\hat{\mathbf{1}}$ and operators generated by $\hat{\Phi}(f)$ with its non-linear 
combinations. It has a structure of a unital $*$-algebra, where the $*$-operation is an involution mapping an element of
$\mathcal{A}_{\mathcal{M}}$ into another its element.\footnote{If one specifies a particular Hilbert space representation
of the algebra, the involution is then identified with the Hermitian conjugation.} Quantum states are defined as linear,
positive and normalized functionals on $\mathcal{A}_{\mathcal{M}}$. A basic reference is~\cite{Haag} and recent reviews~\cite{Hollands&Wald,Khavkine&Moretti,Fewster&Verch} devoted to local quantum field theory in curved spacetimes. 

To fix definitions and notations used below, we will briefly review certain aspects of a (linear) quantum
scalar field in the background of the eternal Schwarzschild--AdS geometry.

The eternal Schwarzschild--AdS black hole is obtained through the Kruskal extension of the 
Schwarzschild black hole in AdS space. We will denote this geometry as $\mathcal{K}$. 
A part of it corresponds to $\mathcal{M}_2'{\cup}\mathcal{M}_2$, 
where $\mathcal{M}_2'$ is a causal complement of $\mathcal{M}_2$. In other words, it is the ``left" outside region of
the black hole, whereas $\mathcal{M}_2$ is the ``right" outside region of the hole (see fig.~1).
One may then choose a subset
of the test functions with vanishing support in either $\mathcal{M}_2$ or $\mathcal{M}_2'$. This leads
to a splitting of the operator algebra $\mathcal{A}_{\mathcal{K}}$ into a product of two factor subalgebras, i.e.
$\mathcal{A}_{\mathcal{M}_2'}{\otimes}\mathcal{A}_{\mathcal{M}_2}$. This procedure is analogous to 
that in the eternal Schwarzschild--Minkowski geometry~\cite{Israel,Sewell,Kay} and~\cite{Wald}.

A physical Hilbert space representation, $\mathcal{H}$, of the algebra $\mathcal{A}_{\mathcal{K}}$ is constructed on the 
Hartle-Hawking state $|\Omega_2\rangle$. This representation is in particular characterised by splitting the field operator 
$\hat{\Phi}(x)$ into a sum of two operators, i.e.
\beqa
\hat{\Phi}(x) &=& \hat{a}(x) + \hat{a}^{\dagger}(x)\,,
\eeqa
where $\hat{a}(h)$ and $\hat{a}^{\dagger}(h)$ are interpreted as the Hartle-Hawking particle annihilation and creation 
operators  corresponding to a wave packet $h(x)$ being a solution of a scalar field equation, where 
\beqa
\hat{a}(h) &\equiv& i\int_{\Sigma} d\Sigma_{\mu}\sqrt{-g(x)}\,g^{\mu\nu}(x)\big(
h^*(x)\nabla_{\nu}\hat{\Phi}(x) - \hat{\Phi}(x)\nabla_{\nu}h^*(x)\big)\,,
\eeqa
where $\Sigma$ is a Cauchy surface, and star denotes a complex conjugation. The wave packet $h(x)$ when 
restricted to the black-hole horizons is positive frequency with respect to $\partial_{U}$ and 
$\partial_V$, where $U$ and $V$ are the Kruskal coordinates in which the horizons are at $U = 0$ and 
$V = 0$. 

On the other hand, one may choose another wave packet $h(x)$ to be positive frequency one with respect to $\partial_u$ 
and $\partial_v$, where $u$ and $v$ are the retarded and advanced Schwarzschild coordinates. One can thus also write
\beqa
\hat{\Phi}(x) &=& \hat{b}_\text{L}(x) + \hat{b}_\text{R}(x) + \text{H.c.}\,,
\eeqa
where the operator $\hat{b}_\text{R}(x)$ and its Hermitian conjugate have zero support in $\mathcal{M}_2'$, while 
$\hat{b}_\text{L}(x)$ and its Hermitian conjugate in $\mathcal{M}_2$. In other words, one rewrites $\mathcal{A}_{\mathcal{K}}$ 
as a product of two commuting sets of field operators $\mathcal{A}_{\mathcal{M}_2'}$ and $\mathcal{A}_{\mathcal{M}_2}$ 
generated by ``left" and ``right" operators, respectively. One may further define two states $|\Omega_\text{L}\rangle$ and 
$|\Omega_\text{R}\rangle$ generating $\mathcal{H}_\text{L}$ and $\mathcal{H}_\text{R}$ and being vacua for $\hat{b}_\text{L}(x)$ 
and $\hat{b}_\text{R}(x)$, respectively. These states are the ``left" and ``right" Boulware vacua. Thus, we introduce a Hilbert space
representation of the algebra on a factorized product, i.e. $\mathcal{H}_\text{L}{\otimes}\mathcal{H}_\text{R}$. 

The Hilbert spaces $\mathcal{H}$ and $\mathcal{H}_\text{L}{\otimes}\mathcal{H}_\text{R}$ cannot be unitarily equivalent in LQFT:
\beqa\label{eq:unitary-inequivalence}
\mathcal{H} &\not\cong& \mathcal{H}_\text{L}{\otimes}\mathcal{H}_\text{R}
\quad \text{or} \quad \mathcal{H} \;\perp\; \mathcal{H}_\text{L}{\otimes}\mathcal{H}_\text{R}\,.
\eeqa
This can be explained in terms of the type III property of factor subalgebras $\mathcal{A}_{\mathcal{M}_2'}$ and 
$\mathcal{A}_{\mathcal{M}_2}$ of $\mathcal{A}_{\mathcal{K}}$ (see, for instance,~\cite{Yngvason}). It essentially means
that the representation $\mathcal{H}$ of $\mathcal{A}_{\mathcal{K}}$ does not factorize into the product 
$\mathcal{H}_\text{L}{\otimes}\mathcal{H}_\text{R}$ under the factorization of $\mathcal{A}_{\mathcal{K}}$ into the
factor subalgebras $\mathcal{A}_{\mathcal{M}_2'}$ and $\mathcal{A}_{\mathcal{M}_2}$. This property has been recently
emphasized in~\cite{Giddings} in applications to the entanglement in LQFT.
\begin{figure}[t]
\includegraphics[width=12.0cm]{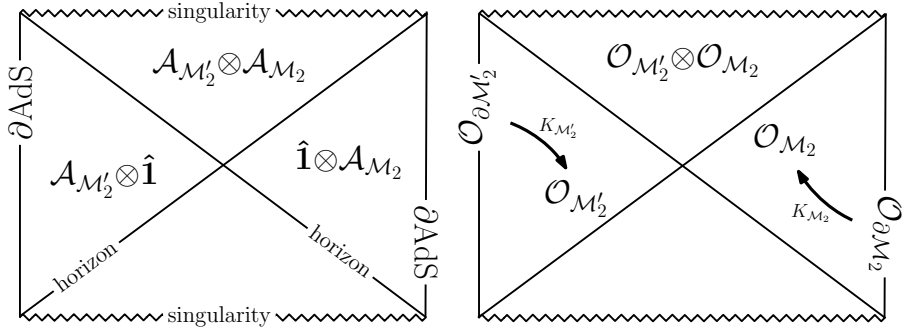}
\caption{A sketch of the Kruskal extension of the Schwarzschild-AdS black hole.}
\end{figure}

The type III property of those factor subalgebras can be demonstrated as follows.
Slightly generalising~\cite{Israel} to the asymptotically AdS case, one may rewrite the Hartle-Hawking 
state as a thermal-field double state~\cite{Umezawa}, i.e.
\beqa\label{eq:tfds}
|\Omega_2\rangle &=& \frac{1}{Z^{\frac{1}{2}}}\prod_{\omega lm}\,
\sum\limits_{n = 0}^{+\infty}\,e^{-\beta E_{\omega,n}/2}\,|n_\text{L}\rangle{\otimes}|n_\text{R}\rangle\,,
\quad\text{where} \quad E_{\omega,n} \;\;\equiv\;\; \omega\,n
\eeqa
and $Z$ is a normalization factor, such that $\langle\Omega_2|\Omega_2\rangle = 1$. The frequency $\omega > 0$ is defined 
with respect to the Killing vector $K_2$, and $l,m$ are the orbital and magnetic numbers referring to a particular representation 
of the rotational symmetry of the black hole. The inverse temperature $\beta$ is given by $1/T_\text{HP}$, where $T_\text{HP}$ 
is the Hawking-Page (HP) temperature~\cite{Hawking&Page}. The states entering the right-hand side of \eqref{eq:tfds} are
defined as
\beqa
|n_\text{L}\rangle &\equiv& \frac{1}{\sqrt{n!}}\,
\big(\hat{b}_{\text{L},\omega lm}^{\dagger}\big)^n|\Omega_\text{L}\rangle\,,
\eeqa
and the same for $|n_\text{R}\rangle$ with $\text{L} \rightarrow \text{R}$ in the above formula. It is worth noting that
$n_{\text{L},\text{R}}$ depend on $\omega$, $l$ and $m$. The state on the left-hand side of \eqref{eq:tfds} is the only 
non-singular state on the black-hole horizons. The normalisation factor $Z$ is infinite (see, for instance,~\cite{Emelyanov5}).
Thus, $|\Omega_2\rangle \not\in \mathcal{H}_\text{L}{\otimes}\mathcal{H}_\text{R}$. Hence, the equation~\eqref{eq:tfds}
is merely formal. This confirms~\eqref{eq:unitary-inequivalence}.

To summarize, any operator in $\mathcal{K}$ can be constructed by employing elements of the algebra
$\mathcal{A}_\mathcal{K} \equiv \mathcal{A}_{\mathcal{M}_2^\prime}{\otimes}\mathcal{A}_{\mathcal{M}_2}$.
There are two ``natural" representations of $\mathcal{A}_\mathcal{K}$, namely $\mathcal{H}$ and
$\mathcal{H}_\text{L}{\otimes}\mathcal{H}_\text{R}$. The former is built on the Hartle-Hawking vacuum $|\Omega_2\rangle$, whereas the latter on the Boulware vacua, i.e. $|\Omega_\text{L}\rangle{\otimes}|\Omega_\text{R}\rangle$. These are unitarily inequivalent (there is no unitary 
operator mapping these into each other), i.e. $\mathcal{H} \not\cong \mathcal{H}_\text{L}{\otimes}\mathcal{H}_\text{R}$.

\subsection{Two copies of CFT and their representations}
\label{subsec:2CFT}

Following the proposal of~\cite{Balasubramanian&Kraus&Lawrence&Trivedi,Maldacena}, physics in the eternal Schwarzschild--AdS geometry is described by two non-interacting (independent) CFT algebras represented on a factorised product of two CFT Hilbert spaces on each AdS boundary. Thus, the physical degrees of freedom are supported on two causally separated boundaries. These degrees of freedom 
are represented by rays in CFT Hilbert spaces. These will be denoted as $\partial\mathcal{H}_\text{L}$ and 
$\partial\mathcal{H}_\text{R}$. Two CFT states corresponding to these Hilbert spaces will be denoted as 
$|\partial\Omega_\text{L}\rangle$ and $|\partial\Omega_\text{R}\rangle$, respectively. At the semi-classical approximation, one actually deals with a thermal gas of excitations belonging to the Hilbert space $\partial\mathcal{H}_\text{L}{\otimes}\partial\mathcal{H}_\text{R}$ 
at the Hawking-Page temperature (see, for instance,~\cite{Horowitz&Polchinski}).

One may expect that the bulk CFT algebra represented in $\partial\mathcal{H}_\text{L}{\otimes}\partial\mathcal{H}_\text{R}$ reduces to local quantum field theory in the large $N$ limit, where $N$ refers to a degree of the gauge group characterising conformal field theory 
on the AdS boundary. This is a sort of the Bohr principle stating that a more fundamental theory (quantum gravity) contains a less fundamental one (semi-classical quantum field theory) in a certain limit ($N \rightarrow \infty$). According to AdS/CFT extended to 
the eternal AdS black hole, one has $\mathcal{H}_{\text{L},\text{R}} \cong \partial\mathcal{H}_{\text{L},\text{R}}$. Since
$\mathcal{H} \not\cong \mathcal{H}_\text{L}{\otimes}\mathcal{H}_\text{R}$, the bulk CFT algebra $\mathcal{O}_{\mathcal{M}_2}$
represented on $\partial\mathcal{H}_\text{R}$ is not unitarily equivalent to the local algebra $\mathcal{A}_{\mathcal{M}_2}$ 
representation on the Hartle-Hawking state. 

An entangled state considered in~\cite{Balasubramanian&Kraus&Lawrence&Trivedi,Maldacena} and leading to the Hilbert space 
$\partial\mathcal{H}$ is presumably identified with the Hartle-Hawking vacuum, such that $\mathcal{H} \cong \partial\mathcal{H}$.
At least, it seems to be a motivation for selecting this state. In other words, one implicitly assumes 
by that the representations of the field operators $\mathcal{A}_{\mathcal{M}_2}$ and $\mathcal{O}_{\mathcal{M}_2}$ are unitarily equivalent 
in the semi-classical limit.\footnote{Note that the algebraic isomorphism between operator algebras 
$\mathcal{A}_{\mathcal{M}_2}$ and $\mathcal{O}_{\mathcal{M}_2}$ is here assumed. This issue will be investigated in detail below.} 

To summarize, the representation of $\mathcal{O}_\mathcal{K} \equiv \mathcal{O}_{\mathcal{M}_2^\prime}{\otimes}\mathcal{O}_{\mathcal{M}_2}$ in $\mathcal{K}$ must be $\partial\mathcal{H}$ which cannot be
understood as a tensor product $\partial\mathcal{H}_\text{L}{\otimes}\partial\mathcal{H}_\text{R}$ of two
(ordinary) CFT Hilbert spaces. In other
words, these are unitarily inequivalent. The eternal-black-hole nucleation is a non-unitary process
$\partial\mathcal{H}_\text{L}{\otimes}\partial\mathcal{H}_\text{R} \rightarrow \partial\mathcal{H}$. Usually,
this kind of dynamics corresponds to phase transitions. This fits well to the idea of having the
Hawking-Page phase transition.

\subsection{Single CFT and its representations}
\label{subsec:1CFT}

Recently there was pronounced a proposal to describe the eternal Schwarzschild--AdS geometry by a single CFT~\cite{Roy&Sarkar}. 
Its possible realisation for different types of AdS black holes is demonstrated below in this section. We will show that a description by
a single CFT is always implicitly present, although one introduces two copies of CFT. This observation follows from a state chosen for 
the CFT operators~\cite{Balasubramanian&Kraus&Lawrence&Trivedi,Maldacena}.

\subparagraph{BTZ black hole.}

To gain insight of how that can be done, let us consider the topological black hole in three-dimensional AdS space, i.e. the 
Ba$\tilde{\text{n}}$ados-Teitelboim-Zanelli (BTZ) black hole~\cite{Banados&Teitelboim&Zanelli&Henneaux} (see~\cite{Carlip1} for a review).\footnote{There is apparently a misuse of terminology in the literature in referring to the BTZ black hole as the eternal Schwarzschild--AdS one. To avoid any possible confusions, these two types of the black holes are distinguished throughout the paper.} 

The BTZ black hole is actually described by a single CFT algebra defined on the boundary of AdS$_3$. To see this one needs first to imagine that the CFT algebra in AdS$_3$ space is split in two factor subalgebras with supports on the ``left" and ``right" AdS-Rindler patches 
of the AdS$_3$ boundary, respectively. A CFT state to be non-singular on the Rindler horizons is the ordinary CFT vacuum (see fig.~2). One then takes the spatial Rindler coordinate be periodic. By doing this, one constructs the BTZ black hole. Thus, the physical Hilbert space representation of the single CFT algebra is constructed on the ordinary CFT vacuum (up to a quotient~\cite{Horowitz&Marolf})\footnote{A certain quotient of AdS hyperboloid corresponding to an evolving universe was also discussed in~\cite{Horowitz&Marolf}. A quantum field theory in this background has been studied in~\cite{Emelyanov}.} and generating the Hilbert space representation $\partial\mathcal{H}$.

Hence, two CFT algebras defined on each boundary of the BTZ black hole are factors of the single CFT algebra 
defined on the AdS$_3$ boundary (up to the quotient). It is analogous to the splitting of $\mathcal{A}_{\mathcal{K}}$ into 
$\mathcal{A}_{\mathcal{M}_2'}{\otimes}\mathcal{A}_{\mathcal{M}_2}$ discussed in Sec.~\ref{subsec:star-algebra}. 

A problem of no holographic description of the BTZ black hole pointed out in~\cite{Avery&Chowdhury} appears only 
if one chooses an improper CFT algebra representation, i.e. $\partial\mathcal{H}_\text{L}{\otimes}\partial\mathcal{H}_\text{R}$,
This representation of the field operators implies the absence of non-local vacuum correlations between casually
unrelated regions (e.g., no Reeh-Schlieder property). 
As discussed above, the quantum gravity degrees of freedom must be associated with rays in $\partial{\mathcal{H}}$, 
rather than in that factorised product.

\subparagraph{Schwarzschild--AdS black hole.}

The AdS black-hole nucleation is a quantum gravity process of the Hawking-Page phase transition from AdS space filled with a 
thermal gas 
to the eternal AdS black-hole geometry~\cite{Hawking&Page}. Let us imagine a flat space composed of two boundaries of the AdS 
black hole, i.e. $\partial\mathcal{K} \equiv \partial\mathcal{M}_2'{\cup}\partial\mathcal{M}_2$, each of which is three-dimensional compactified Minkowski spacetime. The Schwarzschild time translation operator $K_2 = \partial_{t_\text{S}}$ can be employed to set frequency modes in $\partial\mathcal{M}_2'{\cup}\partial\mathcal{M}_2$. However, there are two possible unitarily inequivalent choices 
of the Hilbert space representation: either one considers a representation $\partial\mathcal{H}$ built on $|\partial\Omega_2\rangle$ or 
a representation $\partial\mathcal{H}_\text{L}{\otimes}\partial\mathcal{H}_\text{R}$ defined with respect to
$|\partial\Omega_\text{L}\rangle{\otimes}|\partial\Omega_\text{R}\rangle$.

Each of $\partial\mathcal{M}_2'$ and $\partial\mathcal{M}_2$ should be the whole three-dimensional Minkowski manifold. This can be shown at least in two ways. First, the boundary of AdS space, $\partial\mathcal{M}_1$ ($\equiv \partial\mathcal{M}$), is compactified Minkowski space. The boundary of the Schwarzschild--AdS black hole, $\partial\mathcal{M}_2$, must fit $\partial\mathcal{M}$ to have 
the Hawking-Page transition. Second, the conformal Killing algebra of AdS$_4$ space is given by $\mathfrak{so}(2,4)$. Its generators 
are $L_{\mu\nu}$, $L_{\mu 4}$ and $L_{\mu 5}$, where $\mu,\nu \in \{0,1,2,3\}$. Among of these vectors, $L_{\mu 5} = \{L_{05}\,,L_{i5}\}$ and $L_{\mu\nu} = \{L_{ij}\,,L_{0i}\}$ are generators of the AdS isometry algebra, where $i,j$ run from 1 to 3. The Schwarzschild--AdS geometry has the four-dimensional Killing algebra, such that its Killing vector $K_2$ commutes with the generators of the rotations. The vectors 
$L_{05}$ and $L_{ij}$ are the only commuting generators of $\mathfrak{so}(2,3)$. Thus, the Schwarzschild time translation generator $K_2$ must be identified with the Killing vector $L_{05}$ of the AdS hyperboloid.\footnote{The vector $K_2$ is sometimes called as a boost generator. It is misleading according to the above discussion. However, the boost generator sets indeed dynamics outside of 
the BTZ black hole.} That is $K_2 = K_1$, where $K_1$ is the global AdS time translation operator. This was pointed out 
in~\cite{Banks&Douglas&Horowitz&Martinec}. Hence, $\partial\mathcal{M}_2$ is Minkowski space, i.e. $\partial\mathcal{M}$, in the 
so-called closed coordinates. In these coordinates a spatial topology of Minkowski spacetime is $\mathbf{S}^2$.
\begin{figure}[t]
\includegraphics[width=12.0cm]{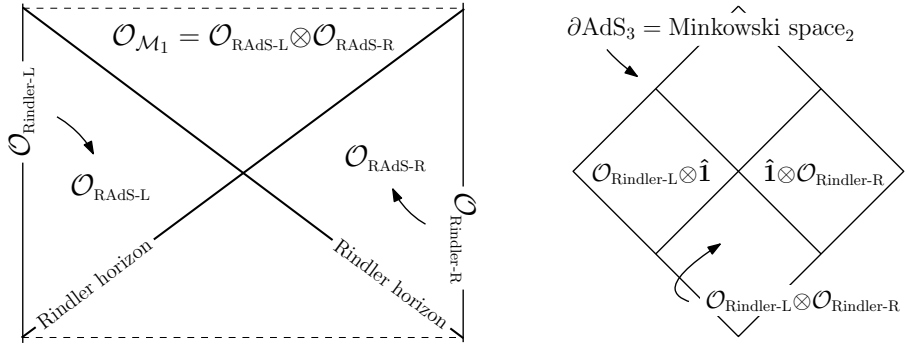}
\caption{``Left" and ``right" AdS-Rindler (RAdS) patch in $\text{AdS}_3$ and their boundary $\partial\text{AdS}_3$. 
The CFT algebra $\mathcal{O}_{\partial\text{AdS}}$ can be represented as a product of two factor subalgebras, i.e.
$\mathcal{O}_{\small\text{Rindler-L}}{\otimes}\mathcal{O}_{\small\text{Rindler-R}}$. The CFT or Minkowski vacuum
probed by either $\mathcal{O}_{\small\text{Rindler-L}}$ or $\mathcal{O}_{\small\text{Rindler-R}}$ responses as a thermal
state, i.e. the vacuum two-point function for each of the factor subalgebras satisfies the Kubo-Martin-Schwinger condition with
respect to the boost generators of the conformal group.}
\end{figure}

If one deals with the pure AdS geometry, then the Hilbert space representation of both theories, i.e.
$\mathcal{A}_{\mathcal{M}_1}$ and $\mathcal{O}_{\mathcal{M}_1}$, are algebraically and unitarily isomorphic at the 
semi-classical approximation~\cite{Rehren} (see~\cite{Kay&Larkin} for its classical counterpart). If one 
further accepts an idea of the eternal black-hole nucleation as pasting together two AdS spaces~\cite{Raamsdonk}, say, 
$\mathcal{M}_1'$ and $\mathcal{M}_1$, then one should associate $\partial\mathcal{H}_\text{L}$ with the theory on 
$\partial\mathcal{M}'$, and $\partial\mathcal{H}_\text{R}$ with the theory on $\partial\mathcal{M}$. It is worth emphasising
that $\partial\mathcal{H}_{\text{L},\text{R}}$ are two CFT Hilbert spaces constructed on the ordinary CFT vacuum. In other
words, the states $|\partial\Omega_\text{L}\rangle$ and $|\partial\Omega_\text{R}\rangle$ are two ordinary CFT vacua in
this subsection. The thermal gas in each AdS space, i.e. $\mathcal{M}_1'$ or $\mathcal{M}_1$, can then be obtained by
considering a Kubo-Martin-Schwinger (KMS) or thermal state defined with respect to $L_{05}$ ($=K_1 = K_2$) in each space.\footnote{The KMS state $|\Omega_{\beta}\rangle$ is a state which satisfies the KMS condition: $\langle\Omega_{\beta}|\alpha_K^t(\hat{A})\hat{B}|\Omega_{\beta}\rangle = 
\langle\Omega_{\beta}|\hat{B}\alpha_K^{t+i\beta}(\hat{A})|\Omega_{\beta}\rangle$, where both sides are analytic in the strip 
$0 < \text{Im}(t) < \beta$, continuous on its boundary, and $\alpha_K^t(\hat{A}) \equiv \exp(+i\hat{K}t)\hat{A}\exp(-i\hat{K}t)$,
$\hat{K}$ is a Hermitian operator corresponding to the Killing vector $K$. More details can be found, for instance,
in~\cite{Haag}.} Thus, the thermal gas is composed of excitations identified with particle states in 
$\partial\mathcal{H}_\text{L}{\otimes}\partial\mathcal{H}_\text{R}$ (see fig.~3).

However, if the black hole has appeared through the Hawking-Page transition, then the Hilbert space of the total system 
should be $\partial\mathcal{H}$ which, as argued above, is not a factorized product of $\partial\mathcal{H}_\text{L}$
and $\partial\mathcal{H}_\text{R}$. The state $|\partial\Omega_2\rangle$ on $\partial\mathcal{M}$ is a pure 
state. It responses as a thermal state for the CFT operators having zero support on 
$\partial\mathcal{M}'$.\footnote{This can be also phrased as follows: $|\partial\Omega_2\rangle$ restricted
to $\partial\mathcal{M}$ (or $\partial\mathcal{M}'$) is a KMS state with respect to the Killing vector $L_{05}$.}
The same occurs for the CFT operators on $\partial\mathcal{M}'$ which have zero support on 
$\partial\mathcal{M}$.

Thus, the Hawking-Page phase transition should be understood as the evolution of the CFT operator 
algebra representation from $\partial\mathcal{H}_\text{L}{\otimes}\partial\mathcal{H}_\text{R}$ to
$\partial\mathcal{H}$. Note that for local quantum field theory characterized by $\mathcal{A}_{\mathcal{M}}$,
one thus has its representation $\mathcal{H}_\text{L}{\otimes}\mathcal{H}_\text{R}$ in 
$\mathcal{M} = \mathcal{M}_1'{\cup}\mathcal{M}_1$ and $\mathcal{H}$ in 
$\mathcal{M} = \mathcal{M}_2'{\cup}\mathcal{M}_2$, respectively, where $\mathcal{H}_{\text{L},\text{R}}$ are
here two AdS Hilbert spaces built on the ordinary AdS vacuum, and $\mathcal{H}$ is the Hilbert space defined in
Sec.~\ref{subsec:star-algebra}.
\begin{figure}[t]
\includegraphics[width=12.0cm]{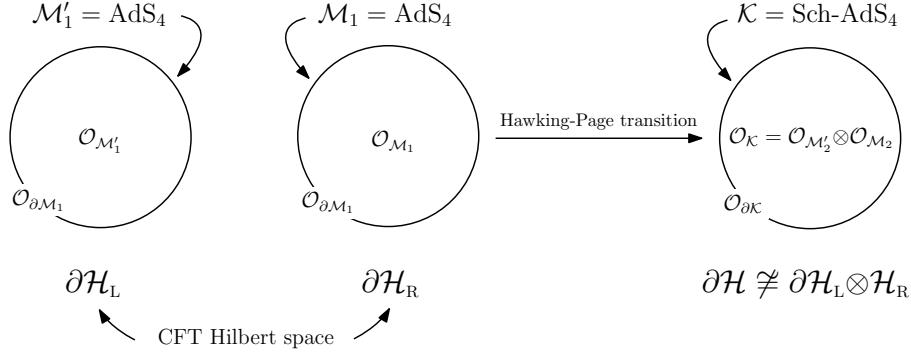}
\caption{A sketch of the Hawking-Page phase transition. Before the transition $\mathcal{M}_1$ is filled by
thermal gas of particles being states in the CFT Hilbert space $\partial\mathcal{H}_R$. If the gas temperature $T$
exceeds the critical temperature $T_c$, then the nucleation of the black hole occurs. This is
described at quantum level by 
$\mathcal{O}_{\mathcal{M}_1^\prime}{\otimes}\mathcal{O}_{\mathcal{M}_1} \rightarrow 
\mathcal{O}_{\mathcal{M}_2^\prime}{\otimes}\mathcal{O}_{\mathcal{M}_2}$ and
$\partial\mathcal{H}_\text{L}{\otimes}\partial\mathcal{H}_R \rightarrow \partial\mathcal{H}$. The Hilbert space
$\partial\mathcal{H}$ is defined with respect to the vacuum $|\partial\Omega_2\rangle$ which cannot be
understood in quantum field theory as an entangled state in $\partial\mathcal{H}_\text{L}{\otimes}\partial\mathcal{H}_R$.
However, it is an (maximally) entangled state in the sense it is not a product state for non-trivial field operators
being products of elements from $\mathcal{O}_{\mathcal{M}_2^\prime}$ and $\mathcal{O}_{\mathcal{M}_2}$.
In other words, the state $|\partial\Omega_2\rangle$ carries non-local (vacuum) correlations. This is in a complete
analogy with the property of the CFT vacuum which carries non-local correlations between the ``left" and ``right"
Rindler patch.}
\end{figure}

To make the above statements more clear, one should restrict the consideration, for instance, to $\partial\mathcal{M}$. 
The physical degrees of freedom could be represented by rays either in $\partial\mathcal{H}_\text{R}$ or $\partial\mathcal{H}$ 
in this boundary. The Hilbert space $\partial\mathcal{H}_\text{R}$ is the Gelfand-Naimark-Segal representation built on 
$|\partial\Omega_\text{R}\rangle$, while $\partial\mathcal{H}$ on $|\partial\Omega_2\rangle$. The thermal gas in 
$\partial\mathcal{M}$ corresponds to a ``thermalized" version of $|\partial\Omega_\text{R}\rangle$ and
described by a thermal density matrix $\hat{\rho}_\text{R}$. However, the pure state 
$|\partial\Omega_2\rangle$ responses as a thermal state merely when probed by operators 
$\mathcal{O}_{\partial\mathcal{K}}$ restricted to $\partial\mathcal{M}$ (or having zero support in 
$\partial\mathcal{M}'$), i.e. $\mathcal{O}_{\partial\mathcal{M}}$ (or $\mathcal{O}_{\partial\mathcal{M}'}$). 
Only in the latter case one has the Tomita--Takesaki construction (e.g., see~\cite{Haag}) which 
is broadly exploited to describe physics inside of the AdS black hole by the CFT operators (see below).

Moreover, one should consider each of $\partial\mathcal{M}'$ and $\partial\mathcal{M}$ as two Lorentzian patches of 
complexified Minkowski spacetime. This can be understood in the case of the spatially flat Minkowski coordinates as follows.
The transition from the ``right" Minkowski space to the ``left" Minkowski space is achieved through $t \rightarrow - t - i\beta/2$,
where $t$ is the Minkowski time coordinate for which its spatial topology is $\mathbf{R}^2$. Indeed, the line element  
$ds^2 = dt^2 - dr^2 - r^2d\theta^2$ can be written as $e^{-2\kappa r}(dT^2-dR^2) - r^2d\theta^2$, where 
$T = e^{\kappa r}\sinh\kappa t$ and $R = e^{\kappa r}\cosh\kappa t > 0$. 
One can thus analytically extend Minkowski space by allowing $R < 0$. It is achieved through
$t \rightarrow - t - i\beta/2 = -t - i\pi/\kappa$.
More concretely, the Minkowski time coordinate has an imaginary part lying in a circle $\mathbf{S}$ with a 
circumference $\beta$. One may further imagine Euclidean Minkowski space, i.e. $t = i\tau$, 
where the Euclidean time $\tau \in \mathbf{S}$. One then considers an Euclidean wave functional 
of a non-interacting CFT model, i.e.
\beqa\label{eq:ewf}
\Psi[\mathcal{O}(\mathbf{y})] &=& \int\mathcal{D}\mathcal{O}\,
\exp\left(-\int_{\tau_\text{R}}^{\tau_\text{L}}d\tau\int d\mathbf{y}\,\big(
\big(\partial\mathcal{O}\big)^2 + (1/8)R\,\mathcal{O}^2\big)\right),
\eeqa
where $y^{\mu} = (\tau,\mathbf{y}) \in \partial\mathcal{M}$, and $\tau_\text{L} = \tau_\text{R} + \beta/2$. 
The wave functional can then be rewritten as
\beqa
\Psi[\mathcal{O}(\mathbf{y})] &=& \frac{1}{Z^{\frac{1}{2}}}\prod_{\mathbf{k}}\,
\sum\limits_{n = 0}^{+\infty}\,e^{-\beta E_{\omega,n}/2}\,
\Psi_{\omega,n}[\mathcal{O}_\text{L}(\mathbf{y})]{\times}\Psi_{\omega,n}[\mathcal{O}_\text{R}(\mathbf{y})]\,.
\eeqa
where $\omega = |\mathbf{k}|$, $\mathbf{k} = (k_x,k_y,k_z)$, 
$\mathcal{O}_{\text{L},\text{R}}(\mathbf{y}) = \mathcal{O}(y)|_{\tau_{\text{L},\text{R}}}$
and $\Psi_{\omega,n}[\mathcal{O}_{\text{L},\text{R}}(\mathbf{y})] = 
\langle\mathcal{O}_{\text{L},\text{R}}(\mathbf{y})|\partial n_{\text{L},\text{R}}\rangle$. The ``left" and ``right" quantities
refer to ``left" ($R < 0$) and ``right" ($R > 0$) Minkowski spaces, respectively. The connection between the path integral
in non-trivial Euclidean backgrounds with the thermal-field dynamics was pointed out in~\cite{Laflamme}.

Analogous formula probably exists if one introduces the so-called closed coordinates in which a spatial topology of Minkowski space 
is $\mathbf{S}^2$ (after conformally mapping it to the closed Einstein static universe). The momentum $\mathbf{k}$ is then substituted by 
$\omega_{\tilde{n}} = 2\tilde{n} + l + \Delta$, $l$ and $m$, where $\tilde{n} \in \mathbf{N}_0$ and $\Delta$ is a conformal weight of 
$\hat{\mathcal{O}}(y)$. However, the mapping between spaces is given by $t_\text{S} \rightarrow - t_\text{S} - i\beta/2$ in this case.
It is worth noting that $t \neq t_\text{S}$.

To sum it up, the algebras $\mathcal{O}_{\partial\mathcal{M}'}$ and $\mathcal{O}_{\partial\mathcal{M}}$ should be
treated as subalgebras of a larger one, i.e.
$\mathcal{O}_{\partial\mathcal{K}} = \mathcal{O}_{\partial\mathcal{M}'}{\otimes}\mathcal{O}_{\partial\mathcal{M}}$.
In other words, these two CFT algebras should be understood as its factor subalgebras. 
The algebras $\mathcal{O}_{\partial\mathcal{M}'}$ and $\mathcal{O}_{\partial\mathcal{M}}$ represented on
the product Hilbert space $\partial\mathcal{H}_\text{L}{\otimes}\partial\mathcal{H}_\text{R}$ built on
$|\partial\Omega_\text{L}\rangle{\otimes}|\partial\Omega_\text{R}\rangle$ can be considered as two independent 
copies of CFT, while the representation $\partial\mathcal{H}$ constructed on the pure state $|\partial\Omega_2\rangle$
can be only treated as associated with the single, enlarged CFT algebra $\mathcal{O}_{\partial\mathcal{K}}$. Thus, 
one should describe the bulk degrees of freedom by the single CFT algebra $\mathcal{O}_{\partial\mathcal{K}}$ 
represented on $\partial\mathcal{H}$. The important point is that $\partial\mathcal{H}$ is unitarily inequivalent
to $\partial\mathcal{H}_\text{L}{\otimes}\partial\mathcal{H}_\text{R}$ or, in other words, 
$|\partial\Omega_2\rangle \notin \partial\mathcal{H}_\text{L}{\otimes}\partial\mathcal{H}_\text{R}$.

\section{Algebraic inequivalence of $\mathcal{A}_{\mathcal{M}_2}$ and $\mathcal{O}_{\mathcal{M}_2}$}
\label{sec:ai}

\subsection{Inside of AdS black hole}
\label{subsec:TTT}

A construction of the bulk CFT operators is achieved in practice through the so-called smearing
function $K_{\mathcal{M}}(x,y)$ which is a kernel of an integral transform, i.e.
\beqa
\hat{\mathcal{O}}_{\mathcal{M}}(x) &=& 
\int d^3y\sqrt{-g(y)}\,K_{\mathcal{M}}(x,y)\hat{\mathcal{O}}(y)\,,
\eeqa
where $x \in \mathcal{M}$ and $y \in \partial\mathcal{M}$. The smearing function was explicitly constructed in AdS 
spacetime, i.e. $\mathcal{M} = \mathcal{M}_1$, in~\cite{Bena,Hamilton&Kabat&Lifschytz&Lowe}. The algebraic 
and unitary equivalence of $\mathcal{A}_{\mathcal{M}_1}$ and $\mathcal{O}_{\mathcal{M}_1}$ represented on the
AdS vacuum and the ordinary CFT vacuum, respectively, can already be seen at the level of two-point correlation
functions of each theory. It may also be expected that $\mathcal{A}_{\mathcal{K}}$ and $\mathcal{O}_{\mathcal{K}}$ 
represented on the Hartle-Hawking vacuum $|\Omega_2\rangle$ and the state $|\partial\Omega_2\rangle$, 
respectively, describe equivalent theories.

The Tomita-Takesaki theorem is extensively exploited to describe physics inside the eternal Schwarzschild--AdS 
black hole.\footnote{The theorem states in particular that von Neumann algebra $\mathcal{R}$
(e.g. $\mathcal{A}_{\mathcal{M}_2}$) represented in a cyclic and
separating state $|\Omega\rangle$ (e.g. $|\Omega_2\rangle$) can be mapped to its commutant $\mathcal{R}'$ 
(e.g. $\mathcal{A}_{\mathcal{M}_2'}$) by an anti-unitary
operator $J$. For a full formulation of the theorem, see~\cite{Haag}. A von Neumann algebra is an algebra 
of bounded operators on a Hilbert space which is closed in the weak topology of matrix elements. For example,
operators of the form $\exp(\hat{\Phi}(f))$ are its elements.} This theorem implies that the algebra 
$\mathcal{A}_{\mathcal{M}_2}$ acting on the Hartle-Hawking state should generate a space which is dense in 
$\mathcal{H}$. In other words,
an observer inhabiting a region outside of the black hole should be able to probe the whole Hilbert space
$\mathcal{H}$ with an arbitrary precision. This can be understood as a Reeh-Schlieder property of von Neumann 
algebra $\mathcal{A}_{\mathcal{M}_2}$.
Although this property seems not to be rigorously proven in the AdS-Schwarzschild space yet, it is analogous 
to that in the eternal Schwarzschild 
geometry~\cite{Kay}.\footnote{For establishing the Reeh-Schlieder property for certain subalgebras 
represented in the AdS vacuum in anti-de Sitter space, see~\cite{Buchholz&Summers} and~\cite{Morrison}.}
This turns out to be possible, partially because $\mathcal{A}_{\mathcal{M}_2}$ and $\mathcal{A}_{\mathcal{M}_2'}$
considered as von Neumann algebras can be mapped into each other by an anti-unitary (modular conjugation) 
operator $J$. This operator corresponds to the transformation $t_\text{S} \rightarrow - t_\text{S} - i\beta/2$ in
spacetime $\mathcal{K}$ discussed above, so that
\beqa
J\mathcal{A}_{\mathcal{M}_2}J &=& \mathcal{A}_{\mathcal{M}_2'}
\eeqa
and vise versa. It is worth emphasising that the theorem is realised for the particular state generating the Hilbert space 
$\mathcal{H}$, rather than the factorised product $\mathcal{H}_\text{L}{\otimes}\mathcal{H}_\text{R}$. 

The same construction should be employed on the AdS boundary if one expects the (extrapolate) AdS/CFT 
correspondence in AdS space with a Schwarzschild black hole. The Tomita-Takesaki theorem is applicable if 
one works with the Hilbert space representation $\partial\mathcal{H}$ of $\mathcal{O}_{\partial\mathcal{K}}$ 
constructed with respect to the state $|\partial\Omega_2\rangle$. To construct CFT operators inside of the
black hole, one needs to extend the boundary by introducing extra Minkowski space 
$\partial\tilde{\mathcal{M}}$. This Minkowski space in flat coordinates corresponds to $T > 0$ and $T > R$. 
It is worth noting that $\partial\mathcal{M}'$, $\partial\mathcal{M}$ and 
$\partial\tilde{\mathcal{M}}$ are not separated by any horizons. However, it is not evident whether this extra 
geometrical construction is acceptable. This question would be interesting to investigate in detail.

It is worth noting that $\partial\tilde{\mathcal{M}}$ is a half of the Milne patch of Minkowski space in the
case of the BTZ black hole (up to the quotient). This patch is a part of the whole boundary of AdS$_3$.
Thus, no extra geometrical construction is needed in the BTZ black-hole case.
\begin{figure}[t]
\includegraphics[width=6.0cm]{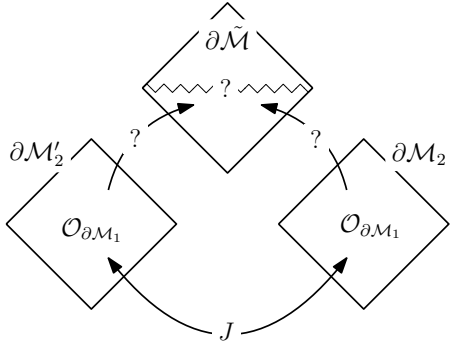}
\caption{A sketch of the boundary of the eternal Schwarschild-AdS black hole.}
\end{figure}

To summarise, it is necessary to understand the boundary structure of the Schwarzschild-AdS black hole
to implement the idea of AdS/CFT in this geometry. This could be a way towards finding a proper mapping
of the CFT operators inside the black hole. 

\subsection{No-Go theorem}
\label{subsec:NGT}

The correspondence principle implies that there is at least algebraic isomorphism between 
$\mathcal{A}_{\mathcal{K}}$ and $\mathcal{O}_{\mathcal{K}}$ in the large $N$ limit. 
To have a unitary isomorphism between them, their Hilbert space representations must by isomorphic (equal up 
to a unitary transform). Both representations are obtained through the Gelfand-Naimark-Segal 
construction~\cite{Haag} built on the pure states $|\Omega_2\rangle$ and $|\partial\Omega_2\rangle$. 
These states are KMS (thermal) ones with respect to the same Killing vector 
$L_{05} \in \mathfrak{so}(2,3)$ and, e.g., subalgebras $\mathcal{A}_{\mathcal{M}_2}$ and
$\mathcal{O}_{\mathcal{M}_2}$, respectively.

It is worth emphasising at this point that the KMS condition depends on two ingredients. These are a field operator
algebra and a one-parameter group of its automorphism. An empty state can also satisfy the KMS condition depending 
on the operator algebra and its automorphism chosen to probe this state. For example, the Minkowski vacuum is a KMS 
state with respect to an operator subalgebra with vanishing support in one of the Rindler patches and its automorphism generated
by the boost Killing vector~\cite{Sewell,Kay}. In the case of a non-interacting, massless scalar field model, the Minkowski vacuum is 
also a (conformal) KMS state with respect to an operator subalgebra with vanishing support either in the contracting or in the expanding
Milne patch and its automorphism generated by the dilatation (see, for instance,~\cite{Emelyanov3}).

Non-uniqueness of a KMS state can appear through the existence of several distinct thermodynamic phases~\cite{Haag}. This can 
be exemplified by the Hawking-Page phenomenon~\cite{Hawking&Page}. Indeed, the same inverse temperature $\beta$ characterises two different states, namely the AdS space with the thermal gas and the Hartle-Hawking state restricted to the region outside of the black hole. In the path integral approach in the saddle-point approximation, both phases are coexisting, described by a non-extremal KMS 
state\footnote{An extremal KMS state is a state that cannot be decomposed into a mixture of other KMS states.} and can be distinguished by an appropriately designed observable~\cite{Emelyanov2}. 

Non-uniqueness of a KMS state can also appear through a spontaneous symmetry breaking~\cite{Haag}. For example, one may 
construct two KMS states with the same inverse temperature under two states one of which is symmetric under $\text{SU}(2)$ and 
another under $\text{U}(1)$ (a phase of the spontaneous magnetisation).

Although both states, the AdS vacuum at the Hawking-Page temperature and the Hartle-Hawking vacuum, are invariant under the 
same subalgebra of $\mathfrak{so}(2,3)$, namely $\mathbf{R}\oplus\mathfrak{so}(3)$, these vacua are different. This originates from different topologies of $\mathcal{M}_1$ and $\mathcal{M}_2$. The topology of the boundary is however unchanged during the Hawking-Page transition. Moreover, it is the same Minkowski space. Therefore, the KMS state defined with respect to $L_{05}$ on the boundaries 
of both $\mathcal{M}_1$ and $\mathcal{M}_2$ is unique, i.e. the thermal two-point~function does not change for
the boundary CFT theory.

A few remarks are in order. First, the effective field theory we are considering in the bulk, i.e. $\mathcal{A}_{\mathcal{M}}$,
is non-interacting. According to the holographic idea, there should exist a certain \emph{free} or \emph{Gaussian} CFT theory on the boundary which may or may not correspond to $\mathcal{A}_{\mathcal{M}}$. In
the case of AdS space, we specified this CFT and denoted it as $\mathcal{O}_{\partial\mathcal{M}}$. Second, the CFT
algebra $\mathcal{O}_{\partial\mathcal{M}}$ is also known as being CFT at infinite $N$~\cite{Kaplan}. One can treat a \emph{generalised free field operator} $\hat{\mathcal{O}}_{\partial\mathcal{M}} \in \mathcal{O}_{\partial\mathcal{M}}$
as composed of some complicated operators of a strongly coupled CFT which is dual to a full UV complete theory
in AdS (e.g., see~\cite{Papadodimas&Raju}). Thus, the uniqueness of the KMS state is related to the operator algebra 
$\mathcal{O}_{\partial\mathcal{M}}$, i.e. the effective non-interacting CFT.

The thermal two-point function can be obtained from the causal propagator~\cite{Haag&Narnhofer&Stein,Narnhofer&Peter&Thirring}:
\beqa
\langle\Omega_{\beta}|\hat{\Phi}_{\mathcal{M}}(x)\hat{\Phi}_{\mathcal{M}}(x')|\Omega_{\beta}\rangle &=&
\frac{i}{2\pi}\int_{\mathbf{R}}\frac{dk}{1 - e^{-\beta k}}\int_{\mathbf{R}}d\delta x^0\,e^{ik\,\delta x^0}
\Delta_{\mathcal{M}}\big(x^0 + \delta x^0,\mathbf{x}|x'\big)\,,
\eeqa
where $\beta$ is the inverse temperature, and $|\Omega_{\beta}\rangle$ is a KMS state defined with respect to a Killing vector 
generating a translation along $x^0$. The causal propagator, i.e. $\Delta_{\mathcal{M}_1}(x,x')$, is given by a difference between 
the retarded and advanced Green function of a field equation satisfied by $\Phi_\mathcal{M}(x)$. This propagator depends on a 
background geometry (stressed out by the lower index $\mathcal{M}$) as well as boundary conditions. In quantum field theory it 
provides the commutator of the field operator at two spacetime points, i.e.
\beqa
i\Delta_{\mathcal{M}_1}(x,x') &=& [\hat{\Phi}_{\mathcal{M}_1}(x),\hat{\Phi}_{\mathcal{M}_1}(x')]
\eeqa
(assuming $\hat{\Phi}(x)$ is non-interacting, i.e. $N$ is infinite). It is worth noting that $\Delta_{\mathcal{M}_1}(x,x')$ is
state-independent. After a certain rescaling of the field operator, $\Delta_{\mathcal{M}_1}(x,x')$ reduces to a CFT causal
propagator in AdS space on its boundary, namely
\beqa\label{eq:cp}
\Delta_{\partial\mathcal{M}}(y,y') &=& \lim\limits_{r \rightarrow \infty}r^{2\Delta}\Delta_{\mathcal{M}_1}(y,r;y',r)\,,
\eeqa
where $\Delta$ is a conformal weight of $\hat{\mathcal{O}}(y)$ ($y \in \partial\text{AdS}$) corresponding to
$\hat{\Phi}_{\mathcal{M}_1}(x)$ in the bulk of AdS, and 
$\Delta_{\partial\mathcal{M}}(y,y') = -i[\hat{\mathcal{O}}(y),\hat{\mathcal{O}}(y')]$. The thermal two-point function of the 
boundary CFT operators depends on a Killing vector chosen to construct this state. The KMS state defined with respect
to $L_{05}$ (the AdS global time translation generator) on the boundary is thus identical to the thermal gas of AdS 
particles in the bulk. In the limit of vanishing temperature, the KMS state thus defined will approach the ordinary AdS
vacuum in the bulk and the ordinary CFT vacuum on the boundary.

During the Hawking-Page transition, the causal propagator of the field $\hat{\Phi}_{\mathcal{M}}(x)$ changes as the
global structure of space changes. In other words, $\Delta_{\mathcal{M}_1}(x,x') \neq \Delta_{\mathcal{M}_2}(x,x')$
as it can be seen, for example, if one compares anti-symmetric combination of two-point correlation functions in
$\mathcal{M}_1$ and $\mathcal{M}_2$ for a scalar field model conformally coupled to gravity~\cite{Emelyanov2}.
However, the boundary is still compactified Minkowski space and the CFT algebra is still invariant under the conformal
algebra $\mathfrak{so}(2,3)$. In other words, $\Delta_{\partial\mathcal{M}}(y,y')$ {\sl cannot} be obtained from
\eqref{eq:cp} with $\Delta_{\mathcal{M}_2}(x,x')$ instead of $\Delta_{\mathcal{M}_1}(x,x')$ on the right-hand side.
Indeed, one has
\beqa
 \lim\limits_{r \rightarrow \infty}r^{2\Delta}\Delta_{\mathcal{M}_1}(y,r;y',r) &\propto& 
\frac{1}{\cos\Delta\Theta - \cos(\Delta{t}_\text{S} - i\varepsilon)} - \frac{1}{\cos\Delta\Theta - \cos(\Delta{t}_\text{S} + i\varepsilon)}\,,
\eeqa 
where $y = (t_\text{S},\theta,\phi)$, $\cos\Delta\Theta \equiv \cos\theta\cos\theta' + \sin\theta\sin\theta'\cos(\phi - \phi')$
and $H_\text{AdS} \equiv 1$, whereas 
\beqa\label{eq:bulk-to-boundary-m2}
 \lim\limits_{r \rightarrow \infty}r^{2\Delta}\Delta_{\mathcal{M}_2}(y,r;y',r) &\propto& \frac{2\Delta\Theta}{\sin\Delta\Theta}\Big(
\frac{1}{\Delta\Theta^2 - (\Delta{t}_\text{S}-i\varepsilon)^2}- \frac{1}{\Delta\Theta^2 - (\Delta{t}_\text{S}+i\varepsilon)^2}\Big)
\eeqa  
in the case, e.g., of the Neumann boundary condition (see~\cite{Emelyanov2} for more details). 

It is usually assumed that it is enough to have a pure high-energy state on the boundary theory to reproduce local
quantum field theory in the bulk of the Schwarzschild--AdS black hole. As it should be evident
from the above discussion, one has at least to supplement that condition by a requirement \eqref{eq:cp} is also
satisfied for $\mathcal{M}_2$. Since this supplementary condition is not realised in the present case, there is no
algebraic isomorphism between the operators $\mathcal{A}_{\mathcal{K}}$ and $\mathcal{O}_{\mathcal{K}}$.

One is hence forced to conclude by a sort of no-go theorem:
\\[2.5mm]
{\sl A description of physics outside/inside of the eternal 
Schwarzschild--AdS black hole is inconsistent with either the
correspondence principle or the holographic principle.}\\[2.5mm]

A few remarks are in order. The holographic principle is here understood in a narrow sense, namely
in terms of the (extrapolate) AdS/CFT idea. Accordingly, the fundamental part of the set-up is the 
conformal, strongly coupled QFT on the boundary. Its subalgebra, i.e. $\mathcal{O}_{\mathcal{M}}$, 
which can be treated as being non-interacting at $N \rightarrow \infty$, should reproduce results
based on an effective local QFT, i.e. $\mathcal{A}_{\mathcal{M}}$, defined on a semi-classical
background $\mathcal{M}$ as expected from the correspondence principle. The consistency of
these principles is well-established in AdS, whereas the modification of the CFT algebra 
$\mathcal{O}_{\mathcal{M}_2}$ is required in Schwarzschild-AdS space as $\mathcal{A}_{\mathcal{M}_2}$
does not acquire conformal symmetry at the \emph{operator} level on the boundary.
One can insist on having $\mathcal{O}_{\mathcal{M}_2}$, but then this theory must be singular near
the black-hole horizon.

\section{Concluding remarks}
\label{sec:discussion}

\subparagraph{CFT Hilbert space representations.}

The first point we would like to emphasiae is that one cannot describe the AdS black hole by two independent CFTs 
if one expects the (extrapolate) AdS/CFT correspondence in the AdS black-hole geometry. This follows from the 
Hilbert space representation that has to be chosen for the CFT algebras on the ``left" and ``right"  eternal AdS 
black-hole boundaries. This representation is $\partial\mathcal{H}$, not the factorised product 
$\partial\mathcal{H}_\text{L}{\otimes}\partial\mathcal{H}_\text{R}$. These CFT algebras, $\mathcal{O}_{\partial\mathcal{M}'}$ 
and $\mathcal{O}_{\partial\mathcal{M}}$, should then be treated as factor subalgebras of the single CFT algebra 
$\mathcal{O}_{\partial{K}}$. It is worth stressing that this construction is state-dependent. Note that it is in agreement 
only to a certain extent with~\cite{Papadodimas&Raju}.

For example, the two CFT algebras on each boundary of the BTZ black hole are factors of a single CFT algebra 
represented on the ordinary CFT vacuum (up to the quotient). The Gelfand-Naimark-Segal representation of the CFT 
algebra on this state is not unitarily equivalent to the factorised product of two CFT Hilbert spaces defined on each 
BTZ boundary. This is in disagreement with~\cite{Balasubramanian&Kraus&Lawrence&Trivedi,Maldacena}, but resolves 
a problem posed in~\cite{Avery&Chowdhury} of no holographic description of the BTZ black hole.

An understanding of the Hawking-Page phase transition as the change of the Hilbert space representation 
of the CFT algebra $\mathcal{O}_{\partial\mathcal{M}'}{\otimes}\mathcal{O}_{\partial\mathcal{M}}$ seems to be
reasonable in the light of~\cite{Witten}. Indeed, the QCD phase transition is characterised by a change of the
representation. This has its imprint in the change of the notion of quasi-particles at the QCD energy scale
$\Lambda_\text{QCD} \approx 0.3\;\text{GeV}$. At low energies, i.e. below $\Lambda_\text{QCD}$, these
particles are 
protons, neutrons, mesons and other hadrons, but at high-energy regime, i.e. at energies well above 
$\Lambda_\text{QCD}$, these are gluons and quarks.

In Sec. \ref{subsec:NGT} we have argued that the KMS state defined w.r.t. $L_{05}$ for the free CFT is unique. The phase transition
$\partial\mathcal{H}_\text{L}{\otimes}\partial\mathcal{H}_\text{R} \rightarrow \partial\mathcal{H}$ does not
change the correlation function of the CFT operator $\hat{\mathcal{O}}(y)$ at two separate boundary points.
This can be understood as follows. Before the phase transition the CFT in the thermal bath of excitations
defined w.r.t. to the CFT vacuum $|\partial\Omega_1\rangle$ and characterised by a density matrix $\hat{\rho}$. 
After the phase transition the CFT is in the pure state $|\partial\Omega_2\rangle$. However, one has
\beqa
\text{Tr}(\hat{\rho}\hat{\mathcal{O}}(y)\hat{\mathcal{O}}(y')) &=& 
\langle\partial\Omega_2|\hat{\mathcal{O}}(y)\hat{\mathcal{O}}(y')|\partial\Omega_2\rangle\,,
\eeqa
where the trace on the right-hand side is taken w.r.t. states of the CFT Hilbert space $\partial\mathcal{H}_R$.
Note that $|\partial\Omega_2\rangle$ provides a vacuum representation of $\mathcal{O}_{\partial\mathcal{K}}$ and is
a KMS state when probed by $\mathcal{O}_{\partial\mathcal{M}}$.

If one rejects the representation evolution, then one could not holographically describe an interior of the black hole. 
This results from the fact that the Tomita-Takesaki theorem broadly exploited to construct CFT operators inside 
the hole is realised only if the CFT algebra is represented in $\partial\mathcal{H}$, not in
$\partial\mathcal{H}_\text{L}{\otimes}\partial\mathcal{H}_\text{R}$. However, this also requires an extra
geometric construction on the boundary (see Sec.~\ref{subsec:TTT}).

\subparagraph{Algebraic inequivalence of $\mathcal{A}_{\mathcal{K}}$ and $\mathcal{O}_{\mathcal{K}}$.}

One may expect that it is enough to choose a ``proper" CFT state defined on the CFT algebra to have a smooth 
horizon in the bulk (see, for instance,~\cite{Papadodimas&Raju}). However, this expectation is 
based on an implicit assumption that $\mathcal{A}_{\mathcal{K}}$ and $\mathcal{O}_{\mathcal{K}}$ are 
algebraically isomorphic in the large $N$ limit. As shown in~\ref{subsec:NGT}, this assumption is invalid.

It is worth noting that this algebraic isomorphism exists in the case of the BTZ black hole. The essential
reason for this is that the BTZ black hole locally corresponds to anti-de Sitter space, i.e. possesses the full
Killing algebra of AdS space.

Presumably, the no-go theorem formulated in Sec.~\ref{subsec:NGT} can be generalised to the Schwarzschild--AdS
black hole formed through the gravitational collapse. It should be possible, because the main argument is based
on comparison the algebraic structure of the semi-classical quantum field operators with that of the CFT ones which 
depend only on the spacetime metric.

It is worth further noting that the high-energy pure state $|E\rangle$ employed in~\cite{Papadodimas&Raju} is 
of different nature in comparison with $|\partial\Omega_2\rangle$. The choice of this high-energy state $|E\rangle$ 
leads merely to the spontaneous AdS symmetry breaking in the large $N$ limit (in the sense of~\cite{Ojima}). 
Modelling a change of spacetime isometry during $\mathcal{M}_1 \rightarrow \mathcal{M}_2$ by the spontaneous 
symmetry breaking on the boundary seems to be impossible, because the algebraic structure of the field operators
depends on a geometry, but not on a state.

Indeed, the state $|E\rangle$ is obtained from the CFT vacuum $|\partial\Omega_1\rangle$ 
by acting on it by a certain CFT operator $\hat{O} \in \mathcal{O}_{\partial\mathcal{M}}$: 
$|E\rangle = \hat{O}|\partial\Omega_1\rangle$. The operator $\hat{O}$ is unitary and well-defined or 
non-anomalous when applied to $|\partial\Omega_1\rangle$. Thus, the Heisenberg and Schr\"{o}dinger
pictures should be equivalent. Therefore, instead of introducing a new state $|E\rangle$ on 
$\mathcal{O}_{\partial\mathcal{M}}$, one can consider a new CFT algebra 
$\mathcal{O}_{\partial\mathcal{M}}^\prime \equiv \hat{O}\,\mathcal{O}_{\partial\mathcal{M}}\,\hat{O}^\dagger$.
The new algebra $\mathcal{O}_{\partial\mathcal{M}}^\prime$ is isomorphic to $\mathcal{O}_{\partial\mathcal{M}}$
even at finite $N$. If one does not take this into account, then one has a pathological theory (e.g., no bulk covariance) 
in the bulk even in AdS space~\cite{Emelyanov6}.

One may consider the CFT bulk operators in the AdS black-hole backgrounds as the only 
available observables. This would be in the spirit of the holography~\cite{tHooft,Susskind}. However,
the bulk CFT operators must be singular on the black-hole horizon, because 
$\mathcal{A}_{\mathcal{K}} \not\cong \mathcal{O}_{\mathcal{K}}$ even in the semi-classical limit. 
This has been actually observed in~\cite{Kabat&Lifschytz}.

\section*{
ACKNOWLEDGMENTS}

It is a pleasure to thank D. Buchholz, C. Germani, S. Konopka, D. Ponomarev, A. Vikman for discussions. 
I am also thankful to R. Verch for discussions and the reference~\cite{Kay}. I am indebted to M. Haack for 
discussions and his valuable comments on an early version of this paper. We are grateful to the anonymous
referee for his comments/questions/suggestions which helped to substantially improve the presentation of
the paper. This research is supported by TRR 33 ``The Dark Universe''.



\begin{thebibliography}{99}

\bibitem{tHooft}
G. t' Hooft,  
\hspace*{0mm}``Dimensional reduction in quantum gravity,''
arXiv:gr-qc/9310026.

\bibitem{Susskind}
L. Susskind,  
\hspace*{0mm}``The world as a hologram,''
J. Math. Phys. {\bf 36}, 6377 (1995)
arXiv:hep-th/9409089.

\bibitem{Maldacena1}
J. Maldacena,
\hspace*{0mm}``The large $N$ limit of superconformal field theories and supergravity,''
Adv. Theor. Math. Phys. {\bf 2}, 231 (1998); arXiv:hep-th/9711200.

\bibitem{Gubser&Klebanov&Polyakov}
S.S. Gubser, I.R. Klebanov, A.M. Polyakov,
\hspace*{0mm}``Gauge theory correlators from non-critical string theory,''
Phys. Lett. B{\bf 428}, 105 (1998); arXiv:hep-th/9802109.

\bibitem{Witten1}
E. Witten,
\hspace*{0mm}``Anti-de Sitter space and holography,''
Adv. Theor. Math. Phys. {\bf 2}, 253 (1998); arXiv:hep-th/9802150.

\bibitem{Balasubramanian&Kraus&Lawrence}
V. Balasubramanian, P. Kraus, A. Lawrence,
\hspace*{0mm}``Bulk versus boundary dynamics in anti-de Sitter spacetime,''
Phys. Rev. D{\bf 59}, 046003 (1999); arXiv:hep-th/9805171.

\bibitem{Banks&Douglas&Horowitz&Martinec}
T. Banks, M.R. Douglas, G.T. Horowitz, E. Martinec,
\hspace*{0mm}``AdS dynamics from conformal field thoery,''
arXiv:hep-th/9808016.

\bibitem{Balasubramanian&Kraus&Lawrence&Trivedi}
V. Balasubramanian, P. Kraus, A. Lawrence, S.P. Trivedi,
\hspace*{0mm}``Holographic probes of anti-de Sitter spacetimes,''
Phys. Rev. D{\bf 59}, 104021 (1999); arXiv:hep-th/9808017.

\bibitem{Giddings1}
S.B. Giddings,
\hspace*{0mm}``(Non)perturbative gravity, nonlocality, and nice slices,''
Phys. Rev. D{\bf 74}, 106009 (2006); arXiv:hep-th/0606146.

\bibitem{Giddings2}
S.B. Giddings,
\hspace*{0mm}``The gravitational S-matrix: Erice lectures,''
Subnucl. Ser. {\bf 48}, 93 (2013); arXiv:hep-th/1105.2036.

\bibitem{Giddings}
S.B. Giddings,
\hspace*{0mm}``Hilbert space structure in quantum gravity: 
an algebraic perspective,'' JHEP12, 099 (2015);
arXiv:hep-th/1503.08207.

\bibitem{Hollands&Wald}
S. Hollands, R.M. Wald,
\hspace*{0mm}``Quantum field theory in curved spacetime,''
arXiv:gr-qc/1401.2026.

\bibitem{Khavkine&Moretti}
I. Khavkine, V. Moretti,
\hspace*{0mm}``Algebraic QFT in curved spacetime and quasi-free Hadamard 
states: an introduction'', arXiv:math-ph/1412.5945.

\bibitem{Fewster&Verch}
Ch.J. Fewster, R. Verch,
\hspace*{0mm}``Algebraic quantum field theory in curved spacetimes,''
arXiv:math-ph/1504.00586.

\bibitem{Bena}
I. Bena,
\hspace*{0mm}``Construction of local fields in the bulk of AdS$_5$ and other spaces,''
Phys. Rev. D{\bf 62}, 066007 (2000), arXiv:hep-th/9905186.

\bibitem{Hamilton&Kabat&Lifschytz&Lowe}
A. Hamilton, D. Kabat, G. Lifschytz, D.A. Lowe,
\hspace*{0mm}``Local bulk operators in AdS/CFT correspondence: a boundary view of horizons and locality,''
Phys. Rev. D{\bf 73}, 086003 (2006), arXiv:hep-th/0506118;
\hspace*{0mm}``Holographic representation of local bulk operators,''
Phys. Rev. D{\bf 74}, 066009 (2006), arXiv:hep-th/0606141;
\hspace*{0mm}``Local bulk operators in AdS/CFT correspondence: a holographic description of the black hole
interior,'' Phys. Rev. D{\bf 75}, 106001 (2007), arXiv:hep-th/0612053.

\bibitem{Rehren}
K.H. Rehren,
\hspace*{0mm}``Local quantum observables in the anti-de Sitter/conformal QFT correspondence,''
Phys. Lett. B{\bf 493}, 383 (2000), arXiv:hep-th/0003120;
\hspace*{0mm}``Algebraic holography,''
Ann. Henri Poincar\'e 1, 607 (2000), arXiv:hep-th/9905179.

\bibitem{Haag}
R. Haag,
\hspace*{0mm}{\sl Local quantum physics. Fields, Particles, Algebras},
(Springer-Verlag, 1996).

\bibitem{Israel}
W. Israel,
\hspace*{0mm}``Thermo-field dynamics of black holes,''
Phys. Lett. A{\bf 57}, 107 (1976).

\bibitem{Sewell}
G.L. Sewell,
\hspace*{0mm}``Quantum fields on manifolds:
PCT and gravitationally induced thermal states,''
Ann. Phys. {141}, 201 (1982).

\bibitem{Kay}
B.S. Kay,
\hspace*{0mm}``A uniqueness result for quasi-free KMS states,''
Helvetica Physica Acta {\bf 58}, 1017 (1985);
\hspace*{0mm}``Purification of KMS states,''
Helvetica Physica Acta {\bf 58}, 1030 (1985);
\hspace*{0mm}``The double-wedge algebra for quantum fields on
Schwarzschild and Minkowski spacetimes,''
Commun. Math. Phys. {\bf 100}, 57 (1985).

\bibitem{Wald}
R.M. Wald,
\hspace*{0mm}{\sl Quantum field theory in curved spacetime and black hole thermodynamics},
(Chicago University Press, 1994).

\bibitem{Yngvason}
J. Yngvason,
\hspace*{0mm}``The role of type $\text{III}$ factors in quantum field theory,''
Rep. Math. Phys. {\bf 55}, 135 (2005); arXiv:math-ph/0411058.

\bibitem{Umezawa}
H. Umezawa,
\hspace*{0mm}{\sl Advanced field theory. Micro, macro and thermal physics},
(AIP Press, 1993).

\bibitem{Hawking&Page}
S.W. Hawking, D.N. Page,
\hspace*{0mm}``Thermodynamics of black holes in anti-de Sitter space,''
Commun. Math. Phys. {\bf 87}, 577 (1983).

\bibitem{Emelyanov5}
S. Emelyanov,
\hspace*{0mm}``Can gravitational collapse and black-hole evaporation be unitary process after all?,''
arXiv:hep-th/1507.03025.

\bibitem{Maldacena}
J. Maldacena,
\hspace*{0mm}``Eternal black holes in anti-de Sitter,''
JHEP04, 021 (2003), arXiv:hep-th/0106112.

\bibitem{Horowitz&Polchinski}
G. Horowitz, J. Polchinski,
\hspace*{0mm} ``Gauge/gravity duality'', (in {\sl Approaches to quantum gravity. 
Toward a new understanding of space, time and matter}, edited by D. Oriti, Cambridge University Press, 2009).

\bibitem{Roy&Sarkar}
S.R. Roy, D. Sarkar,
\hspace*{0mm}``Hologram of a pure state black hole,''
arXiv:hep-th/1505.03895.

\bibitem{Banados&Teitelboim&Zanelli&Henneaux}
(a) M. Ba$\tilde{\text{n}}$ados, C. Teitelboim, J. Zanelli,
\hspace*{0mm}``Black hole in three-dimensional spacetime,''
Phys. Rev. Lett. {\bf 69}, 1849 (1992);
arXiv:hep-th/9204099;
(b) M. Ba$\tilde{\text{n}}$ados, M. Henneaux, C. Teitelboim, J. Zanelli,
\hspace*{0mm}``Geometry of the 2+1 black hole,''
Phys. Rev. D{\bf 48}, 1506 (1993);
arXiv:hep-th/9302012.

\bibitem{Carlip1}
S. Carlip,
\hspace*{0mm}``The (2+1)-dimensional black hole,''
Class. Quant. Grav. {\bf 12}, 2853 (1995); arXiv:gr-qc/9506079.

\bibitem{Horowitz&Marolf}
G.T. Horowitz, D. Marolf,
\hspace*{0mm}``A new approach to string cosmology,''
JHEP07, 014 (1998); arXiv:hep-th/9805207.

\bibitem{Emelyanov}
S. Emelyanov,
\hspace*{0mm}``Freely moving observer in (quasi) anti-de Sitter space,''
Phys. Rev. D{\bf 90}, 044039 (2014), arXiv:gr-qc/1309.3905;
\hspace*{0mm}``Local thermal observables in spatially open FRW spaces,''
Phys. Rev. D{\bf 91}, 124068 (2015), arXiv:gr-qc/1406.3360.

\bibitem{Avery&Chowdhury}
S.G. Avery, B.D. Chowdhury,
\hspace*{0mm}``No holography for eternal AdS black holes,''
arXiv:hep-th/1312.3346.

\bibitem{Kay&Larkin}
B.S. Kay, P. Larkin,
\hspace*{0mm}``Pre-holography,''
Phys. Rev. D{\bf 77}, 121501 (2008);
arXiv:hep-th/0708.1283.

\bibitem{Raamsdonk}
M.Van Raamsdonk,
\hspace*{0mm}``Building up spacetime with quantum entanglement,''
Gen. Rel. Grav. {\bf 42}, 2323 (2010), arXiv:hep-th/1005.3035.

\bibitem{Laflamme}
R. Laflamme,
\hspace*{0mm}``Geometry and thermodynamics,''
Nucl. Phys. B{\bf 324}, 233 (1988).

\bibitem{Buchholz&Summers}
D. Buchholz, S.J. Summers,
\hspace*{0mm}``Stable quantum systems in anti-de Sitter space:
causality, independence, and spectral properties,''
J. Math. Phys. {\bf 45}, 4810 (2004), arXiv:math-ph/0407011.

\bibitem{Morrison}
I.A. Morrison,
\hspace*{0mm}``Boundary-to-bulk maps for AdS causal wedges and
the Reeh-Schlieder property in holography,''
JHEP05, 053 (2014), arXiv:hep-th/1403.3426.

\bibitem{Emelyanov3}
S. Emelyanov,
\hspace*{0mm}``Non-unitarity or hidden observables?,''
arXiv:gr-qc/1410.6149.

\bibitem{Emelyanov2}
S. Emelyanov,
\hspace*{0mm}``Observing quantum gravity in asymptotically AdS space,''
Phys. Rev. D{\bf 92}, 124062 (2015);
arXiv:hep-th/1504.05164.

\bibitem{Kaplan}
J. Kaplan,
\hspace*{0mm}``Lectures on AdS/CFT from the bottom up'' (2015).

\bibitem{Papadodimas&Raju}
K. Papadodimas, S. Raju,
\hspace*{0mm}``An infalling observer in AdS/CFT,''
JHEP10, 212 (2013), arXiv:hep-th/1211.6767;
\hspace*{0mm}``Comments on the Necessity and Implications of State-Dependence in the Black Hole Interior,''
arXiv:hep-th/1503.08825.

\bibitem{Haag&Narnhofer&Stein}
R. Haag, H. Narnhofer, U. Stein,
\hspace*{0mm}``On quantum field theory in gravitational background,''
Commun. Math. Phys. {\bf 94}, 219 (1984).

\bibitem{Narnhofer&Peter&Thirring}
H. Narnhofer, I. Peter, W. Thirring,
\hspace*{0mm}``How hot is the de Sitter space?,''
Int. J. Mod. Phys. B{\bf 10}, 1507 (1996).

\bibitem{Witten}
E. Witten,
\hspace*{0mm}``Anti-de Sitter space, thermal phase transition, and confinement in gauge theories,''
Adv. Theor. Math. Phys. {\bf 2}, 505 (1998), arXiv:hep-th/9803131.

\bibitem{Ojima}
I. Ojima,
\hspace*{0mm}``Lorentz invariance versus temperature in QFT,''
Lett. Math. Phys. {\bf 11}, 73 (1986).

\bibitem{Emelyanov6}
S. Emelyanov,
\hspace*{0mm}``A note on bulk locality and covariance in AdS/CFT ,''
arXiv:hep-th/1507.07897.

\bibitem{Kabat&Lifschytz}
D. Kabat, G. Lifschytz,
\hspace*{0mm}``Finite $N$ and the failure of bulk locality: black holes in AdS/CFT,''
JHEP09, 077 (2014), arXiv:hep-th/1405.6394.
\end{thebibliography}
\end{document}